# Image-based reconstruction for the impact problems by using DPNNs

Yu Li[a, *], Hu Wang[a, **], Wenquan Shuai[a], Honghao Zhang[b, c], Yong Peng[b, c]

*a. State Key Laboratory of Advanced Design and Manufacturing for Vehicle Body, College of Mechanical and Vehicle Engineering, Hunan University, Changsha, 410082, PR China*

*b. Key Laboratory of Traffic Safety on Track of Ministry of Education, School of Traffic and Transportation Engineering, Central South University, Changsha 410000, PR China*

*c. Joint International Research Laboratory of Key Technology for Rail Traffic Safety, Central South University, Changsha, 410000, PR China*

**Abstract**

With the improvement of the pattern recognition and feature extraction of Deep Neural Networks (DPNNs), image-based design and optimization have been widely used in multidisciplinary researches. Recently, a Reconstructive Neural Network (ReConNN) has been proposed to obtain an image-based model from an analysis-based model [1, 2], and a steady-state heat transfer of a heat sink has been successfully reconstructed. Commonly, this method is suitable to handle stable-state problems. However, it has difficulties handling nonlinear transient impact problems, due to the bottlenecks of the Deep Neural Network (DPNN). For example, nonlinear transient problems make it difficult for the Generative Adversarial Network (GAN) to generate various reasonable images. Therefore, in this study, an improved ReConNN method is proposed to address the mentioned weaknesses. Time-dependent ordered images can be generated. Furthermore, the improved method is successfully applied in impact simulation case and engineering experiment. Through the experiments, comparisons and analyses, the improved method is demonstrated to outperform the former one in terms of its accuracy, efficiency and costs.

---

[*] First author. *E-mail address*: liyu_hnu@hnu.edu.cn (Y. Li)

[**] Corresponding author. Tel.: +86 0731 88655012; fax: +86 0731 88822051. *E-mail:* wanghu@hnu.edu.cn (H. Wang)

Time-dependent ordered images.

**Highlights**

i. The nonlinear transient impact problem is successfully reconstructed.

ii. The adversarial algorithm is integrated with the VAE and termed AVAE. It achieves enhanced performance.

iii. The ordered images in term of time can be generated based on the manifold learning model.

iv. A specific CGAN is suggested for image postprocessing.

v. An engineering impact problem of a thin-walled structure is reconstructed well by the EReConNN.

**Nomenclature**

| | | | |
|---|---|---|---|
| DPNN | Deep Neural Network | CNN | Convolutional Neural Network |
| ReConNN | Reconstructive Neural Network | CGAN | Conditional GAN |
| EReConNN | Enhanced ReConNN | MSE | Mean Square Error |
| CWGAN | Compressed Wasserstein GAN | KL | Kullback–Leibler |
| GAN | Generative Adversarial Network | GD | Gradient Descent |
| AE | Autoencoder | JS | Jensen-Shannon |
| VAE | Variational AE | DOF | Degree of Freedom |
| AVAE | Adversarial VAE | SRGAN | Super-Resolution GAN |
| LI | Lagrange Interpolation | ML | Machine Learning |
| DL | Deep Learning | ESPCN | Efficient Sub-Pixel CNN |
| PFHS | Plate Fin Heat Sink | PSNR | Peak Signal-to-Noise Ratio |
| CAD | Computer Aided Design | SSIM | Structural Similarity |
| CIC | Convolution in Convolution | | |
| DRCN | Deep Reconstruction-Classification Network | | |
| SIMP | Solid Isotropic Material with Penalization | | |

## 1. Introduction

Currently, with the explosive development of Machine Learning (ML), some ML-based methods, including the Deep Neural Network (DPNN), which is the core technology of Deep Learning (DL), have been utilized in many interdisciplinary studies, such as computational mechanics [3-10], heat transfer [11-14], fluid mechanics [15-18], etc. Moreover, with the improvements of the pattern recognition and feature extraction of the DPNNs, increasingly more researchers have attempted to solve some engineering problems based on images. E.g., Lin [19], Sosnovik [20], Yu [21] and Banga [22] et al used CNN (Convolutional Neural Network) based models to recognize and extract the features of the initial designs of the topologically optimized designs, and predicted the optimized structure. The testing results based on the Solid Isotropic Material with Penalization (SIMP) method validated that the CNN could significantly reduce the optimization time. For cracks, Fan [23], Dung [24], Dorafshan [25], Cha [26], Chen [27], Yokoyama [28] and Tong [29] et al extracted the features of crack images using DPNNs. Through comparisons between the conventional edge detection and the DPNN methods, the DPNNs outperformed the conventional methods in crack detection. Additionally, with respect to material studies, Li [13] stated that the traditional methods for studying the effective thermal conductivities of composite materials were all based on a good physical understanding, and so he utilized the pattern recognition of the CNN to infer the effective thermal conductivities of composite materials. Cang [30] proposed a generative model that created an arbitrary amount of artificial material samples when trained on only a limited amount of authentic samples. The key contribution of this work was the introduction of a morphology constraint to the training of the generative model. For other fields, Wang [31] proposed a novel full closed-loop approach to detect and classify power quality disturbances based on a deep CNN, and the field data from a multimicro grid system were used to further prove the validity of the proposed method.

Recently, Reconstructive Neural Network (ReConNN) was proposed as a reconstructive model with the distinctive characteristic framework "from analysis-based models to image-based models". It was developed in Ref. [1] and further applied to a heat transfer problem of a 3D Plate Fin Heat Sink (PFHS) in Ref. [2]. The ReConNN model was applied to physical field reconstructions and to construct models that included more objective information. However, considering the accuracy, efficiency and costs, both the ReConNN that proposed in Ref. [1] and the slightly improved version in Ref. [2] were in exploratory stages. Some shortcomings in the existing ReConNN are described and analyzed as follows.

i. The ReConNN was mainly composed of the CNN and Generative Adversarial Network (GAN). The CNN was employed to construct the mapping from the images to the objective functions, while the GAN was utilized to generate more similar images. Nevertheless, because most studies were based on optimization or convergence problems, the CNN might have difficulties constructing high-accuracy mappings from input images to labels[1], and the GAN does not generate various reasonable images.

ii. As shown in Fig. 1 (a), the design domain of the PFHS that was previous solved was fixed during the simulation process. This weak-nonlinear steady-state problem could be handled easily by the ReConNN. However, for nonlinear transient problems, such as the impact problem as shown in Fig. 1 (b), it was powerless.

iii. During the ReConNN, a distinct characteristic of both the CNN and GAN was big data, meaning expensive simulations. However, this study was mainly designed for small samples, and if the necessary dataset was too large, the study would be meaningless.

iv. At the end of the work, an interpolation algorithm was needed to complete the reconstruction. Nonetheless, the matching between the interpolated objective functions and the generated images was inefficient.

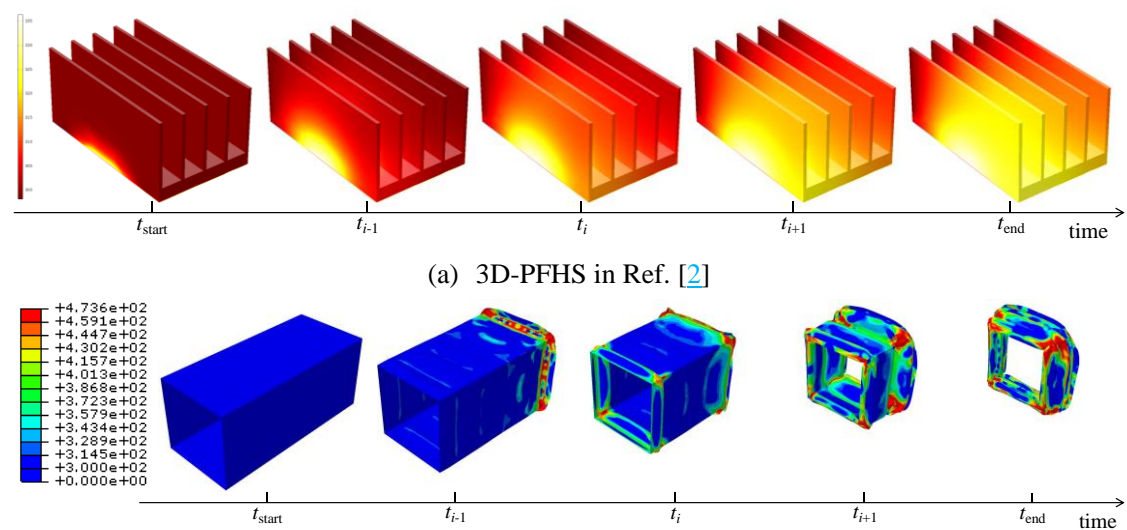

(a) 3D-PFHS in Ref. [2]

---

[1] For NNs, labels denote the real response value of evaluation of samples.

(b) Impact problem in this study
**Fig. 1.** Nonlinearity comparisons between different problems.

To address the above mentioned shortcomings, an Enhanced ReConNN (EReConNN) model is proposed. It abandons the previous integrated architecture of the CNN and the GAN. It uses an Adversarial Variational Autoencoder (AVAE) model to generate the time-dependent ordered images, and avoids the matching work between the generated images and corresponding objective functions in the final reconstruction. Through the tests, comparisons and experiments, the EReConNN outperforms the former one with respect to the accuracy and efficiency, even when nonlinearity exists.

The remainder of this study is organized as follows. In Section 2, the basic idea of the impact problem is introduced. Subsequently, Section 3 is devoted to illustrating the structure of the EReConNN. Meanwhile, the generation process of time-dependent ordered images and the reason why the ReConNN is powerless in this study are also described. Then, some detailed numerical examples, results, and analyses are presented in Section 4. In Section 5, the EReConNN reconstructs an actual engineering problem, which is an impact process of a combined multicell thin-walled aluminum structure. Ultimately, some prospective remarks are provided in the final section.

## 2. Problem descriptions

*2.1. Physical model*

The 3D Computer Aided Design (CAD) model of the impact problem is presented in Fig. 2. The impact body is a cuboid whose material is Al alloy 6,061-T6, as shown in Table 1. It is defined with an initial velocity $v_0$ along the negative direction of the *z*-axis. Furthermore, a 300 *kg* point mass is coupled in the center of the other side of the impact surface. The total impact time is 7 *ms*.

**Table 1** The material parameters of the Al alloy 6,061-T6.

| Parameter | Value |
|---|---|
| Young's modulus (*MPa*) | 71,275 |

| | Yield stress (*MPa*) | Plastic deformation |
|---|---|---|
| Poisson's ratio | 0.33 | |
| Yield stress (*MPa*) | 241.5 | |
| Density (*t/mm³*) | 2.9×10⁻⁹ | |
| | 241.5 | 0 |
| | 263.0 | 0.0069 |
| | 278.8 | 0.0217 |
| | 318.8 | 0.0921 |
| Hardening curve | 346.7 | 0.1408 |
| | 374.5 | 0.1914 |
| | 388.8 | 0.2181 |
| | 423.8 | 0.2862 |
| | 464.3 | 0.3728 |
| | 473.6 | 0.4078 |

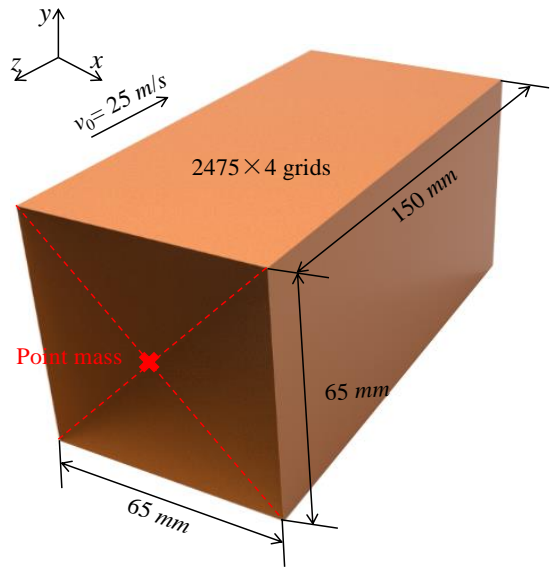

**Fig. 2.** The CAD model of the impact problem.

*2.2. Mathematical model*

In the impact process, the work is done by the external forces during the deformation. If there is no heat loss, the work will be completely converted into strain energy.

$$U = A \tag{1}$$

The strain energy of the per unit volume is calculated by

$$U_0 = U_0\left(\varepsilon_x, \varepsilon_y, \varepsilon_z, \gamma_{xy}, \gamma_{yz}, \gamma_{zx}\right) \tag{2}$$

s.t.

$$\gamma_{xy} = \frac{1}{G}\tau_{xy}, \gamma_{yz} = \frac{1}{G}\tau_{yz}, \gamma_{zx} = \frac{1}{G}\tau_{zx} \tag{3}$$

where

$$G = \frac{E}{2(1+v)} \tag{4}$$

Here, $A$ is the work done by the external forces; $U$ is the strain energy; $\varepsilon$ is the strain; $E$ and $v$ are the Young's modulus and the Poisson's ratio that define the elastic properties, respectively; $\tau$ is the shear stress; and $G$ is the shear modulus.

The distortion and destruction of the impact body generally contain two stages, elastic deformation and plastic deformation. When the body is in the elastic-plastic stage, the stress and strain do not yet correspond to each other. The equation of the strain compatibility of a 3-dimensional elastic-plastic is represented as

$$\begin{cases} \dfrac{\partial^2 \varepsilon_x}{\partial y^2} + \dfrac{\partial^2 \varepsilon_y}{\partial x^2} = \dfrac{\partial^2 \gamma_{xy}}{\partial x \partial y} \\ \dfrac{\partial^2 \varepsilon_y}{\partial z^2} + \dfrac{\partial^2 \varepsilon_z}{\partial y^2} = \dfrac{\partial^2 \gamma_{yz}}{\partial y \partial z} \\ \dfrac{\partial^2 \varepsilon_z}{\partial x^2} + \dfrac{\partial^2 \varepsilon_x}{\partial z^2} = \dfrac{\partial^2 \gamma_{zx}}{\partial z \partial x} \\ 2\dfrac{\partial^2 \varepsilon_x}{\partial y \partial z} = \dfrac{\partial}{\partial x}\left(-\dfrac{\partial \gamma_{yz}}{\partial x} + \dfrac{\partial \gamma_{xz}}{\partial y} + \dfrac{\partial \gamma_{xy}}{\partial z}\right) \\ 2\dfrac{\partial^2 \varepsilon_y}{\partial z \partial x} = \dfrac{\partial}{\partial y}\left(-\dfrac{\partial \gamma_{yz}}{\partial x} + \dfrac{\partial \gamma_{xz}}{\partial y} + \dfrac{\partial \gamma_{xy}}{\partial z}\right) \\ 2\dfrac{\partial^2 \varepsilon_z}{\partial x \partial y} = \dfrac{\partial}{\partial z}\left(-\dfrac{\partial \gamma_{yz}}{\partial x} + \dfrac{\partial \gamma_{xz}}{\partial y} + \dfrac{\partial \gamma_{xy}}{\partial z}\right) \end{cases} \tag{5}$$

The simplest form of linear elasticity is the isotropic case, and the stress-strain relationship is given by

$$\begin{bmatrix} \varepsilon_x \\ \varepsilon_y \\ \varepsilon_z \\ \gamma_{xy} \\ \gamma_{xz} \\ \gamma_{yz} \end{bmatrix} = \begin{bmatrix} 1/E & -v/E & -v/E & 0 & 0 & 0 \\ -v/E & 1/E & -v/E & 0 & 0 & 0 \\ -v/E & -v/E & 1/E & 0 & 0 & 0 \\ 0 & 0 & 0 & 1/G & 0 & 0 \\ 0 & 0 & 0 & 0 & 1/G & 0 \\ 0 & 0 & 0 & 0 & 0 & 1/G \end{bmatrix} \begin{bmatrix} \sigma_x \\ \sigma_y \\ \sigma_z \\ \sigma_{xy} \\ \sigma_{xz} \\ \sigma_{yz} \end{bmatrix} \tag{6}$$

where $\sigma$ is the stress.

In the plastic stage, the characteristics of the stress and strain are nonlinear and not unique, and the strain-state is related to not only the stress but also the stress change. The plastic constitutive relation can be represented as

$$\begin{cases} d\varepsilon_x^p = \dfrac{d\varepsilon_i}{\sigma_i}\left[\sigma_x - \dfrac{1}{2}(\sigma_y+\sigma_z)\right], d\varepsilon_{xy}^p = \dfrac{3}{2}\dfrac{d\varepsilon_i}{\sigma_i}\tau_{xy} \\ d\varepsilon_y^p = \dfrac{d\varepsilon_i}{\sigma_i}\left[\sigma_y - \dfrac{1}{2}(\sigma_x+\sigma_z)\right], d\varepsilon_{yz}^p = \dfrac{3}{2}\dfrac{d\varepsilon_i}{\sigma_i}\tau_{yz} \\ d\varepsilon_z^p = \dfrac{d\varepsilon_i}{\sigma_i}\left[\sigma_z - \dfrac{1}{2}(\sigma_x+\sigma_y)\right], d\varepsilon_{zx}^p = \dfrac{3}{2}\dfrac{d\varepsilon_i}{\sigma_i}\tau_{zx} \end{cases} \tag{7}$$

It can be further shortened as the Levy-Mises function.

$$d\boldsymbol{\varepsilon}_{ij}^p = \dfrac{3}{2}\dfrac{d\varepsilon_i}{\sigma_i}\mathbf{S}_{ij} \tag{8}$$

s.t.

$$\mathbf{S}_{ij} = \begin{bmatrix} \sigma_x - \sigma_m & \tau_{xy} & \tau_{xz} \\ \tau_{yx} & \sigma_y - \sigma_m & \tau_{yz} \\ \tau_{zx} & \tau_{zy} & \sigma_z - \sigma_m \end{bmatrix} \tag{9}$$

where

$$\sigma_m = \sigma_1 + \sigma_2 + \sigma_3 \tag{10}$$

Here, $\varepsilon^p$ is the plastic strain increment; and $\mathbf{S}_{ij}$ is the deflection stress tensor or stress deviator.

*2.3. Implicit algorithm for dynamic problems*

Dynamic problems are mainly researched with respect to the dynamic response of the structure in basic motions or under dynamic forces. Currently, the Newmark-$\beta$ method [32] is one of the most widely used implicit algorithms for dynamical systems with arbitrary excitation.

The initial structural equation of the motion for a linear system with dynamic forces is calculated by

$$M_0\ddot{r}(t) + C_0\dot{r}(t) + K_0 r(t) = F_0(t) \tag{11}$$

where $M_0$, $C_0$, and $K_0$ are the initial mass, damping and stiffness matrices, respectively;

$\ddot{r}(t)$, $\dot{r}(t)$ and $r(t)$ are the functions of time $t$ for the nodal displacement, the nodal velocity and the nodal acceleration, respectively; and $F_0(t)$ is the initial load vector of the nodal force with arbitrary excitation.

To compute the initial structural dynamic response using the Newmark-$\beta$ method, first it needs to determine the $M_0$, $C_0$, and $K_0$. Then, $\dot{r}(t)$ and $r(t)$ can be obtained, and the initial nodal acceleration is represented by

$$\ddot{r}(0) = M_0^{-1}\left[F_0(0) - C_0\dot{r}(t) + K_0 r(0)\right] \tag{12}$$

After that, the following related coefficients can be inferred according to time step $\Delta t$ and parameters $\gamma$ and $\beta$.

$$a_0 = \frac{1}{\beta \Delta t^2}, a_1 = \frac{\gamma}{\beta \Delta t}, a_2 = \frac{1}{\beta \Delta t}, a_3 = \frac{1}{2\beta} - 1,$$

$$a_4 = \frac{\gamma}{\beta} - 1, a_5 = \Delta t\left(\frac{\gamma}{2\beta} - 1\right), a_6 = \Delta t(1-\gamma), a_7 = \gamma \Delta t \tag{13}$$

Finally, the effective stiffness matrix $\tilde{K}_0$ can be presented by

$$\tilde{K}_0 = K_0 + a_0 M_0 + a_1 C_0 \tag{14}$$

As for each time step, the effective load vector $\tilde{F}$ at $t+\Delta t$ is

$$\tilde{F}(t+\Delta t) = F_0(t+\Delta t) + M_0\left[a_0 r(t) + a_2\dot{r}(t) + a_3\ddot{r}(t)\right] + C_0\left[a_1 r(t) + a_4\dot{r}(t) + a_5\ddot{r}(t)\right] \tag{15}$$

The correspondingly nodal displacement $r$, nodal acceleration $\ddot{r}$ and nodal velocity $\dot{r}$ are given by Eqs. (16) - (18).

$$r(t+\Delta t) = \frac{\tilde{F}_0(t+\Delta t)}{\tilde{K}_0} \tag{16}$$

$$\ddot{r}(t+\Delta t) = a_0\left[r(t+\Delta t) - r(t)\right] - a_2\dot{r}(t) - a_3\ddot{r}(t) \tag{17}$$

$$\dot{r}(t+\Delta t) = \dot{r}(t) + a_6\ddot{r}(t) + a_7\ddot{r}(t+\Delta t) \tag{18}$$

## 3. The architecture of the proposed EReConNN

Different from the ReConNN whose main tasks are image regression (Convolution in Convolution, CIC) and image generation (Compressed Wasserstein

GAN, CWGAN), the EReConNN is mainly composed of feature extraction, physical field reconstruction and visualization enhancement by using the AVAE and Conditional GAN (CGAN), respectively.

As shown in Fig. 3, in Step i, the contour image of each iteration during the simulation is collected. In this study, the mappings on *zOy* and *x=y=z* are chosen as the subjects being investigated. Then the AVAE is employed to extract the features of the physical field in Step ii. After that, the time-dependent ordered feature values are interpolated in Step iii. Subsequently, all features are decoded by the decoder of the AVAE that is trained in Step ii and the time-dependent ordered images can be generated. The reconstruction can be completed. Finally, the CGAN is applied to enhance the visualization of the reconstruction.

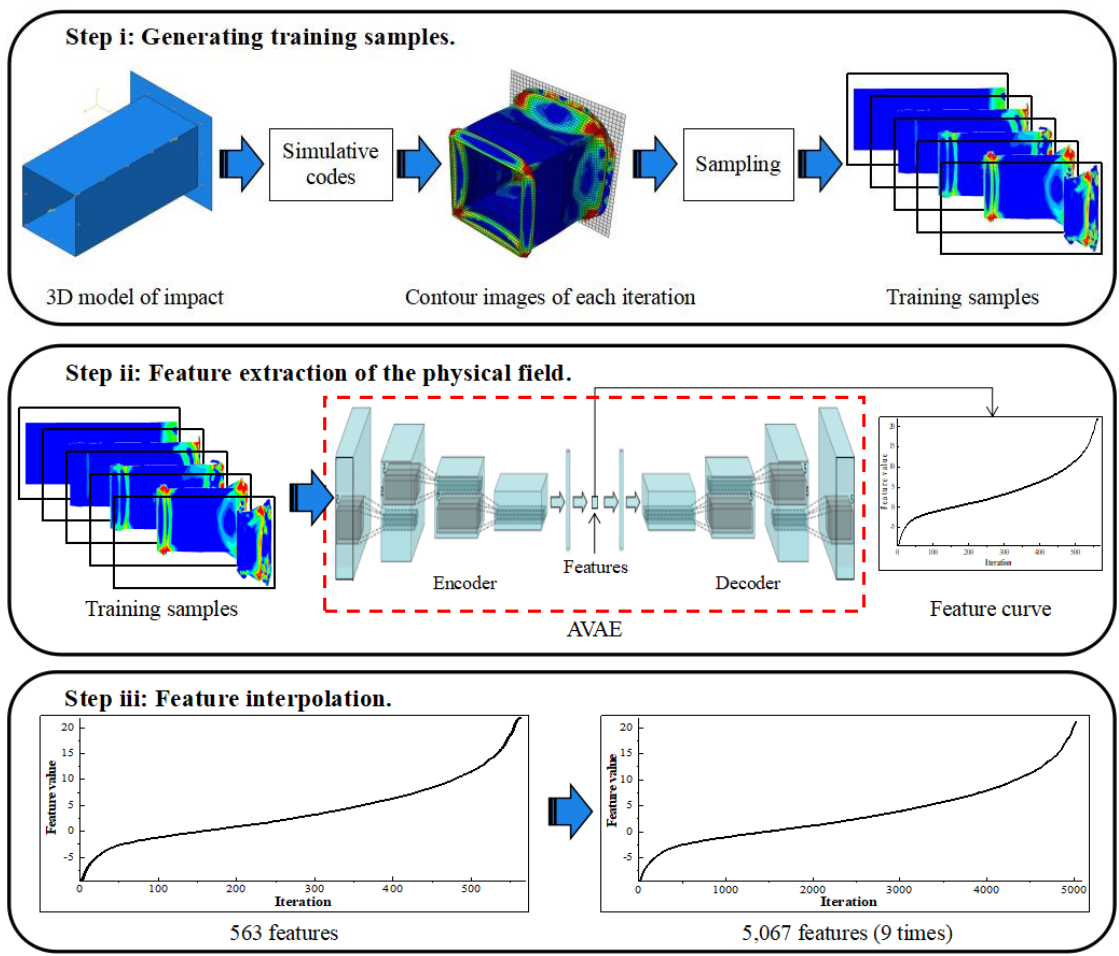

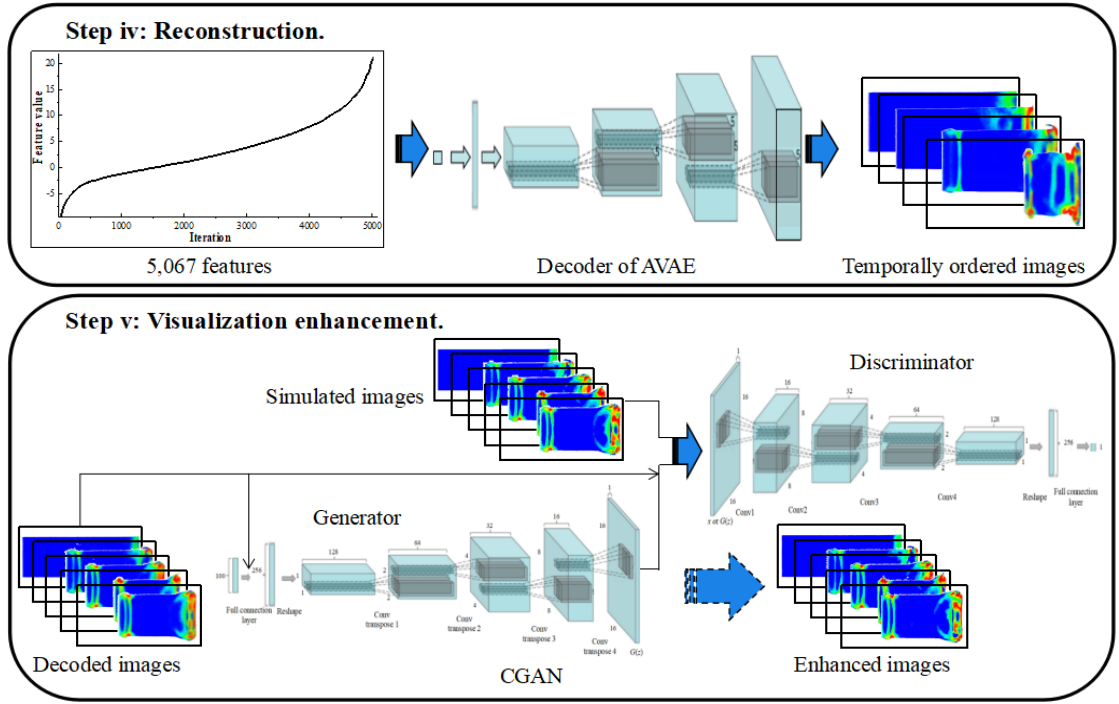

**Fig. 3.** The architecture of the EReConNN model.

*3.1. Bilinear Interpolation*

The pixel sizes of each simulated image are 480×960×3 and 785×880×3 when mapping on the surfaces of *xOy* and *x=y=z*, respectively. To improve the universality of the algorithm, all images are resized to 256×256×3 by using the Bilinear Interpolation before training.

In mathematics, the Bilinear Interpolation is an extension of the Linear Interpolation for interpolating the functions of two variables (e.g., *x* and *y*) on a rectilinear 2D grid. The key idea is to perform linear interpolation first in one direction, and then again in another direction. Although each step is linear, the interpolation as a whole is not linear but rather quadratic.

If a point (*x*, *y*) of an unknown function *f(x)* is interpolated through another four points $Q_{ii}=(x_i, y_i)$, $Q_{ij}=(x_i, y_j)$, $Q_{ji}=(x_j, y_i)$ and $Q_{jj}=(x_j, y_j)$, the linear interpolation in the *x*-direction is first done and this yields

$$f(x, y_i) \approx \frac{x_j - x}{x_j - x_i} f(Q_{ii}) + \frac{x - x_i}{x_j - x_i} f(Q_{ji}) \tag{19}$$

$$f(x, y_j) \approx \frac{x_j - x}{x_j - x_i} f(Q_{ij}) + \frac{x - x_i}{x_j - x_i} f(Q_{jj}) \tag{20}$$

After that, the interpolation in the *y*-direction is processed to obtain the desired estimate.

$$f(x,y) \approx \frac{1}{(x_j - x_i)(y_j - y_i)} \begin{bmatrix} x_j - x & x - x_i \end{bmatrix} \begin{bmatrix} f(Q_{ii}) & f(Q_{ij}) \\ f(Q_{ji}) & f(Q_{jj}) \end{bmatrix} \begin{bmatrix} y_j - y \\ y - y_i \end{bmatrix} \quad (21)$$

*3.2. Adversarial Variational Autoencoder*

An Autoencoder (AE) [33] is a feed-forward NN that is trained to approximate the identity function. That is, it is trained to map from a vector of values to the same vector. When it is used for dimensionality reductions, the first half of the network (encoder) is a model that maps from high-dimensional to low-dimensional spaces, and the second half (decoder) maps from low-dimensional to high-dimensional spaces. Compared with the AE, the outputs from the encoder in the VAE [34, 35] have two purposes: one represents the mean of a Gaussian distribution (z_mean, $\mu$), and the another represents the logarithmic value of the variance of a Gaussian distribution (z_log_var, $\log\sigma^2$). The input to the decoder can be calculated by

$$z = \mu + \varepsilon \cdot \exp\left(\frac{\log\sigma^2}{2}\right) \quad (22)$$

where $\varepsilon \sim N(0, 1)$.

The optimized objective mainly includes two parts. One is the Mean Squared Error (MSE) that is calculated by Eq. (23). The smaller the MSE is, the more similar the predicted values are to the real samples. The other is the Kullback–Leibler (KL) divergence that is represented by Eq. (24). It constrains z_mean and z_log_var. The optimized objective is to minimize the weighted summation of the MSE and the KL divergence, as expressed by Eq. (25).

$$MSE = \frac{1}{n}\sum(y - \hat{y})^2 \quad (23)$$

$$KL = 1 + \log\sigma^2 - \mu^2 - \sigma^2 = 1 + \log\sigma^2 - \mu^2 - \exp(\log\sigma^2) \quad (24)$$

$$L_{VAE} = 0.5 MSE - 0.5 KL \quad (25)$$

where *y* is the training sample; $\hat{y}$ is the predicted value; and *n* is the sample size.

In this study, in order to improve the image restoration and feature extraction abilities, the adversarial algorithm is integrated with the VAE as the AVAE, as shown in Fig. 4. Similar to the GAN, a discriminator is added. Through judging whether the input is real or fake, the images decoded by the VAE can be further improved. As shown in Table 2, the encoder, decoder and discriminator mainly contain 6 convolutional or up sampling layers, respectively. A batch norm layer is added to each convolutional layer to normalize the data and improve the training speed, except for the last convolutional layer in the discriminator. In the encoder, after the convolutional process, a full connection layer is used to obtain the features with the required dimension. Meanwhile, a dropout layer is employed in the decoder and the discriminator to avoid the overfitting problem. The loss of the discriminator is calculated by

$$L_D = \mathbb{E}(D(x)) - \mathbb{E}[D(G(z))] \tag{26}$$

where $x$ is the image from the training samples; and $z$ is a noise vector.

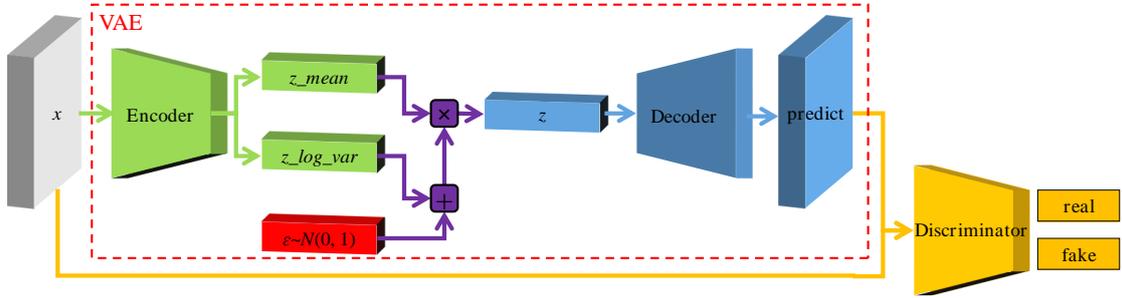

**Fig. 4.** The architecture of the AVAE.

**Table 2** The detailed architecture parameters of the AVAE.

|  | Layer | Kernel size | Kernel stride | Output deep |
|---|---|---|---|---|
| Encoder | Conv-Batchnorm 1 | 4×4 | 2 | 32 |
|  | Conv-Batchnorm 2 | 4×4 | 2 | 32×2 |
|  | Conv-Batchnorm 3 | 4×4 | 2 | 32×4 |
|  | Conv-Batchnorm 4 | 4×4 | 2 | 32×8 |
|  | Conv-Batchnorm 5 | 4×4 | 2 | 32×16 |
|  | Conv-Batchnorm 6 | 4×4 | 2 | 32 |
|  | Full connection | Output nodes = Feature dimension | | |
| Decoder | Full connection | Output nodes = 512 | | |
|  | Up sampling-Dropout 1 | 4×4 | 2 | 32×16 |
| | Up sampling-Dropout 2 | 4×4 | 2 | 32×8 |
| | Up sampling-Dropout 3 | 4×4 | 2 | 32×4 |
| | Up sampling-Dropout 4 | 4×4 | 2 | 32×2 |
| | Up sampling-Dropout 5 | 4×4 | 2 | 32 |
| | Up sampling | 4×4 | 2 | 3 |
| | Conv-Batchnorm 1 | 4×4 | 2 | 32 |
| | Conv-Batchnorm 2 | 4×4 | 2 | 32×2 |
| Discriminator | Conv-Batchnorm 3 | 4×4 | 2 | 32×4 |
| | Conv-Batchnorm 4 | 4×4 | 2 | 32×8 |
| | Conv-Batchnorm 5 | 4×4 | 2 | 32×16 |
| | Conv-Dropout | 4×4 | 2 | 1 |

*3.2.1. The analyses of the powerless of the CWGAN in the ReConNN.*

Compared with the EReConNN, the generative model of the ReConNN is the CWGAN. Meanwhile, through tests, the CWGAN might be powerless for a nonlinear case in this study. The main reasons for the unsatisfactory results are discussed as follows.

· Firstly, as mentioned by Goodfellow, training GANs requires finding a Nash equilibrium [36]. However, only when the function is convex can the Gradient Descent (GD) algorithm realize the Nash equilibrium, which means that the GAN has difficulties reaching the Nash equilibrium in every training.

· The GAN is defined as a min-max problem without loss functions, and so it is difficult to determine if the direction of the training process is right.

· Importantly, compared with other generative models, e.g., AVAE, the GAN uses a Gaussian or uniform distribution to approximate the real data. However, if the input image is too large, too many pixels make the GAN uncontrollable. This is why the images were compressed using a VAE.

· Nevertheless, although the compressed images decrease, the input to the GAN is changed from image data to discrete data. In the GAN, the output from the generator passes a Softmax layer [37] using Eq. (27). The Softmax regression changes the input to a probability distribution. Meanwhile, the final output will be a one-hot matrix. In ML, a one-hot is a group of bits in which the legal combinations of values are only those with a single high (one) bit and all the others low (zero) [38]. However, the discriminator might give the same judgement for different inputs, which causes the GAN to be unsuitable for learning discrete data. For example, the one-hot matrices of (0.2, 0.3, 0.1, 0.2) and (0.2, 0.25, 0.2, 0.1) are both (0, 1, 0, 0). Furthermore, the Jensen-Shannon (JS) divergence [39] in Eqs. (28) and (29) is used as the training objective of the GAN, and it is also inappropriate for addressing discrete data. Despite the WGAN replacing the JS divergence with the Wasserstein distance [40] in Eq. (30), the ability to learn discrete data is still limited.

$$softmax(y_i) = y_i' = \frac{e^{yi}}{\sum_{i=1}^{n} e^{yi}} \tag{27}$$

$$JS(P\|Q) = \frac{1}{2} KL\left(P(x) \Big\| \frac{P(x)+Q(x)}{2}\right) + \frac{1}{2} KL\left(Q(x) \Big\| \frac{P(x)+Q(x)}{2}\right) \tag{28}$$

s.t.

$$KL(P\|Q) = \sum P(x) \log \frac{P(x)}{Q(x)} \tag{29}$$

$$W(P_1, P_2) = \inf_{\gamma \sim \Gamma(P_1, P_2)} \mathbb{E}_{(x,y) \sim \gamma}\left[\|x - y\|\right] \tag{30}$$

where $[y_1, y_2, …, y_i]$ is an input tensor to the Softmax layer; and $\Gamma(P_1, P_2)$ denotes the collection of all measures with marginals $P_1$ and $P_2$ on the first and second factors, respectively.

*3.2.2. The analyses of the powerful of the AVAE in the EReConNN.*

Essentially, the AVAE is a manifold learning model. There are two main distinguishing features of the manifold learning, one is the nonlinear dimensionality reduction, and the other describes the data characterization.

As for the dimensionality reduction, high-dimensional data, meaning data containing more than two or three dimensions, can be difficult to interpret. One approach to simplification is to assume that the data of interest lie on an embedded nonlinear manifold within the lower-dimensional space. If the manifold has a low enough dimension, the data can be visualized.

The data characterization reflects the things that can represent the essential data. Manifold learning "remembers" the data by "learning" the data characterization, which is similar to a human brain.

These two features can help the AVAE outperform other methods in this study. As shown in Fig. 5, each image is regarded as a data point and each pixel is a dimension. Therefore, an image is an $m \times n$-dimensional point in the Euclidean space. If the manifold has a low enough dimension, the images can be distributed in a 1-dimensional space. In addition, if the manifold is learned by the AVAE, the linear relation in the local points of the low-dimensional space is as same as the one in the high-dimensional space. Namely, both have

$$x = w_1 x_1 + w_2 x_2 + \ldots + w_i x_i \tag{31}$$

Thus, through trimming (interpolation) the data in the low-dimensional manifold space, meaningful and reasonable data in the high-dimensional space can be obtained.

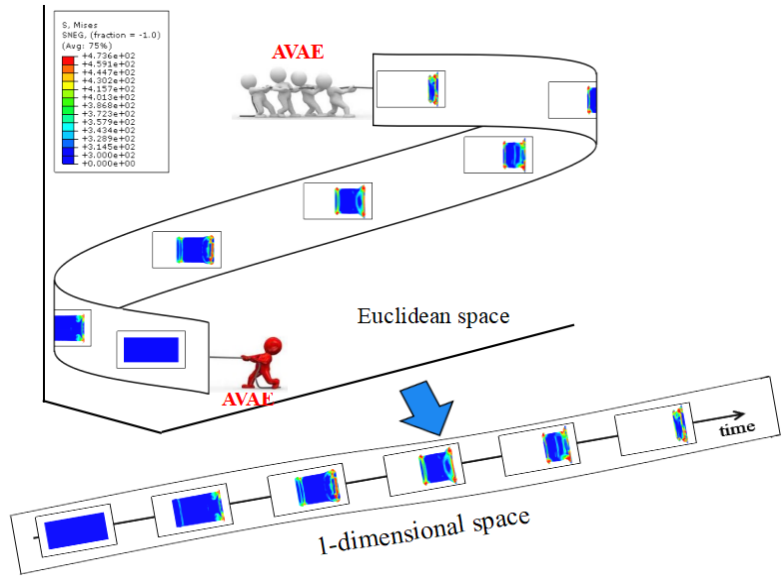

**Fig. 5.** The diagrammatic sketch of the manifold learning.

By the way, the GAN has difficulties inferring a pixel using another pixel, and it can only generate all pixels at the same time. In contrast, as for the AVAE, the images in the manifold space are relevant to each other, and so it is possible to generalize new examples using interpolation. Herein, through the interpolation of the extracted features, the new images in the high-dimensional manifold space can be generated. Thus, the amount of code and the computational costs can be largely saved.

*3.2.3. Generation of time-dependent ordered images.*

As shown in Fig. 6, disordered samples are used to train the AVAE. The AVAE can be regarded as two models: one is feature extraction model that includes the encoder, and the other is the image generation model that includes the decoder. After training, the features of the time-dependent ordered images are extracted using the trained encoder, and it can be seen that each dimension of the features can be distributed in a smooth curve. Then, an interpolation algorithm is employed to obtain more features, and the corresponding time-dependent ordered images can be finally obtained by the trained decoder.

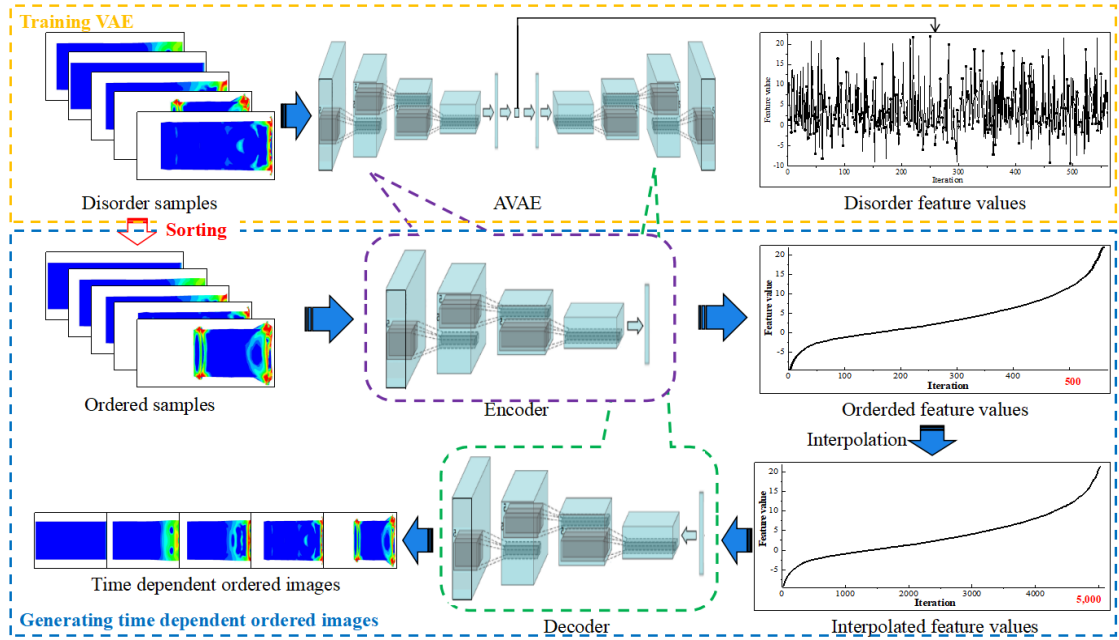

F**ig. 6.** Generation of time-dependent ordered images.

*3.3. Conditional Generative Adversarial Network*

The disadvantage of small samples for the AVAE is the low accuracy, which makes it necessary to do image postprocessing. Currently, there are many methods to improve images' resolution, such as the Super-Resolution GAN (SRGAN) [41], the SRCNN [42], the Deep Reconstruction-Classification Network (DRCN) [43], the Efficient Sub-Pixel CNN (ESPCN) [44], etc. However, these methods may not be suitable for this problem because they aim to reconstruct high-resolution images based on low-resolution images, e.g., the SRGAN increases the resolution by 4 times.

Meanwhile, the purpose of this section is to make the generated images more similar to the actual simulated results without changing resolutions.

In this study, the CGAN is employed to enhance the visualization. As shown in Fig. 7, the GAN [45] is a powerful generative model. It is a method for learning a data distribution $P_{model}(x)$ and realizing a model to sample from it. The GAN consists of two functions: the generator $G(z)$ that maps a sample depending on a random or a Gaussian distribution, and the discriminator $D(x)$ that determines if an input belongs to the training data set. $G(z)$ and $D(x)$ are typically learned jointly by alternate training based on game theory principles. Mathematically, the training process can be described as

$$\min_G \max_D V(D,G) = \mathbb{E}_{x \sim Pdata}\left[\log D(x)\right] + \mathbb{E}_{z \sim Pz}\left[\log\left(1 - D(G(z))\right)\right] \quad (32)$$

where $x$ is the image from training samples $P_{data}$; and $z$ is a noise vector that is sampled from the distribution $P_z$.

As shown in Fig. 7, the CGAN [46] is an extension of GANs where both $G(z)$ and $D(x)$ receive an additional conditioning variable $c$, yielding $G(z, c)$ and $D(x, c)$, respectively. This formulation allows $G(z)$ to generate images that are conditioned on $c$. $c$ can be based on multiple information, e.g., classification labels [47], partial data for image restoration [48] or data from different modalities [46]. Mathematically, the optimized objective of the CGAN is

$$\min_G \max_D V(D,G) = \mathbb{E}_{x \sim Pdata}\left[\log D(x|c)\right] + \mathbb{E}_{z \sim Pz}\left[\log\left(1 - D(G(z|c))\right)\right] \quad (33)$$

Additionally, the specific structure of the CGAN in this study is shown in Fig. 8. It can be seen that the condition is set as the input to both the generator and the discriminator. As for the generator, it looks like an AE. Firstly, the condition is convoluted, and the convoluted result of each layer will be input to the corresponding upsampling layer. Thus, the generator can generate a fake image based on the condition. The contact layer contacts the two tensors in the deep direction of the images, namely, the final dimension of the data. The structure of the discriminator is similar to the convolutional architecture of the generator.

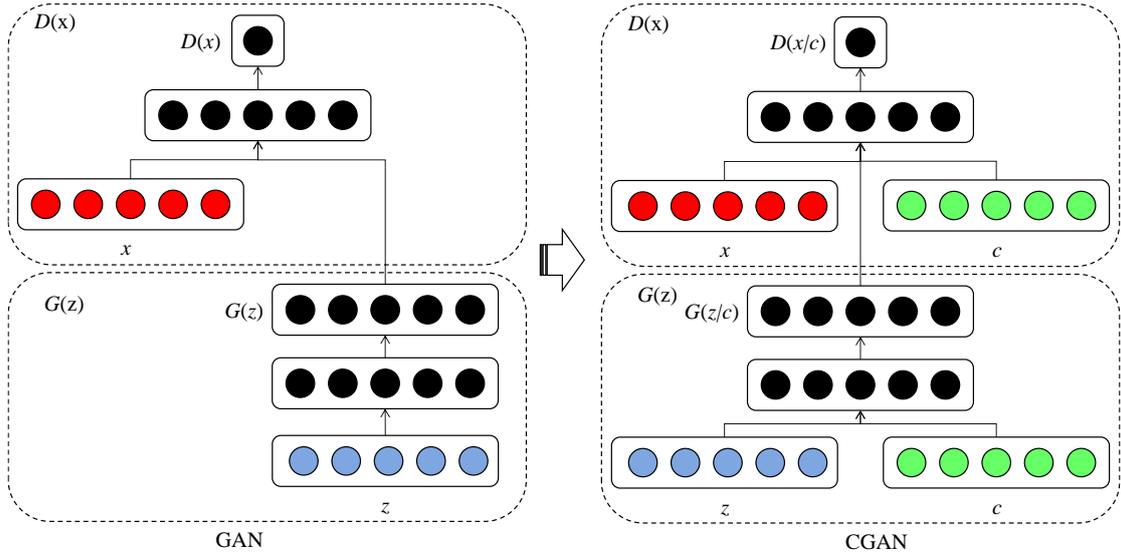

**Fig. 7.** The architectures of the GAN and the CGAN.

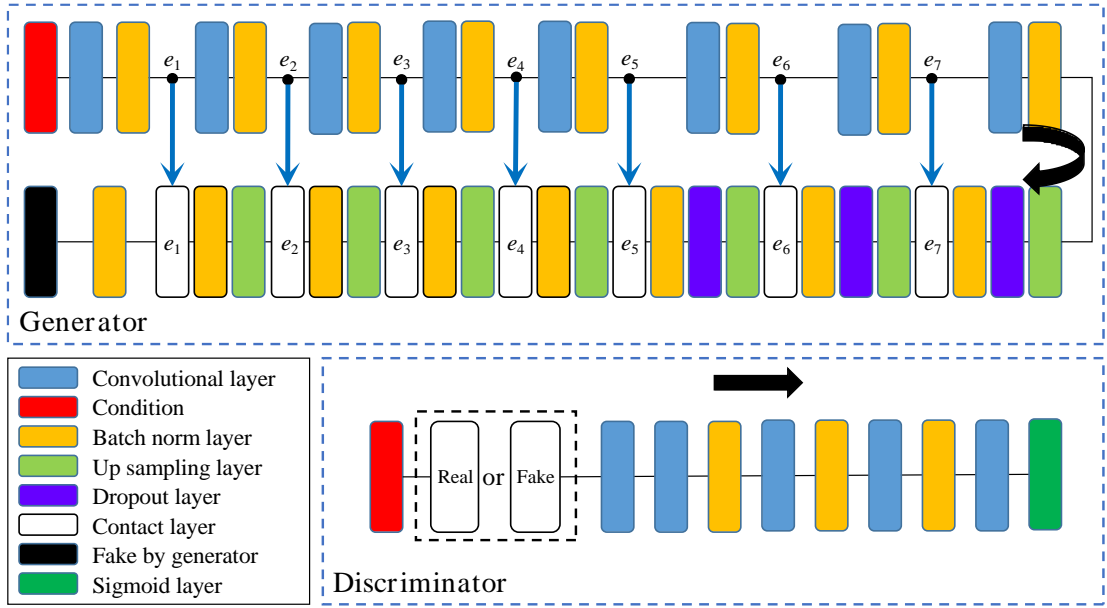

**Fig. 8.** The specific structure of the CGAN used in this study.

## 4. Tests and analyses

In order to evaluate the performance of the EReConNN, experiments, comparisons and analyses are presented in this section. To comprehensively represent the reconstruction, two mappings on the surfaces of *xOy* and *x=y=z* are reconstructed, respectively.

*4.1. The comparisons between AVAE and VAE*

To compare the AVAE and VAE, the Peak Signal-to-Noise Ratio (PSNR) [49]

and Structural Similarity (SSIM) [50] are employed. They are calculated by Eqs. (34) - (35) and Eq. (36), respectively. The PSNR is an engineering term for the ratio between the maximum possible power of a signal and the power of the corrupting noise that affects the fidelity of its representation. When given a noise-free $m \times n$ monochromic image $I$ and its noisy approximation $K$, Eq. (23) can be changed to

$$MSE = \frac{1}{nm}\sum_{i=0}^{n-1}\sum_{j=0}^{m-1}\left[I(i,j) - K(i,j)\right]^2 \qquad (34)$$

Then, the PSNR is defined as

$$PSNR = 10 \cdot \log_{10}\left(\frac{MAX^2}{MSE}\right) = 20 \cdot \log_{10}(MAX) - 10 \cdot \log_{10}(MSE) \quad (\text{dB}) \qquad (35)$$

where $MAX$ is the maximum possible pixel value of the image. Usually, it is 255.

As for the SSIM, it is used to measure the similarity between two images. The SSIM is designed to improve traditional methods such as the PSNR and MSE. It is a value between 0 and 1, and the larger the SSIM is, the better the image quality. It is calculated by

$$SSIM = \frac{(2\mu_I\mu_K + c_1)(2\sigma_{IK} + c_2)}{(\mu_I^2 + \mu_K^2 + c_1)(\sigma_I^2 + \sigma_K^2 + c_2)} \qquad (36)$$

such that

$$c_1 = (k_1 L)^2, c_2 = (k_2 L)^2 \qquad (37)$$

where $\mu_I$ and $\mu_K$ are the average pixels of images $I$ and $K$, respectively; $\sigma_I^2$ and $\sigma_K^2$ are the pixel variances of images $I$ and $K$, respectively; $\sigma_{IK}$ is the covariance of $I$ and $K$; $c_1$ and $c_2$ stabilize the division with weak denominators; and $L$ is the dynamic range of the pixel values. $L=255$, $k_1=0.01$ and $k_2=0.03$ by default.

In this section, a case sample that is the same as the one in Section 4.2 with 563 is employed. The feature dimension is set as 1. After 150 training epochs[2], the mean values of the predicted results are shown in Table 3. The MSE of the AVAE is reduced by approximately 47% compared with the VAE. For the PSNR and SSIM, the AVAE has remarkable increases.

---

[2] In one epoch, all samples should have been trained once.

Table 3 Predicted results by using the AVAE and VAE.

| Method | PSNR | SSIM | MSE |
|---|---|---|---|
| AVAE | 21.08dB | 89.21% | 5.50E-3 |
| VAE | 19.93dB | 84.64% | 1.03E-2 |

*4.2. The reconstruction of the mapping on the xOy surface*

The mapping on the *xOy* surface of the 3D impact problem looks like a 2D case. From this mapping surface, the computational ability requirements of the AVAE and the CGAN are relatively low.

*4.2.1. Feature extraction of the physical field.*

In this section, the dimension of the nonlinear transient case is reduced and its essential features will be extracted. The first and most important issue is to ensure suitable dimensions for the features. As shown in Fig. 9, with the convolutional layers, the linear features of the physical field are gradually extracted to map the physical field from high-dimensional to low-dimensional spaces. In this section, the dimensions of the features are set as 4, 64, 256 and 784, and the average VAE losses of the total data set are shown in Fig. 10. Each dimensional case is trained for 844,950 steps, namely, 150 epochs, and the number of simulation iterations is 5,633[3]. It can be seen that when the dimension is less than 256, the loss values are not very different. Although it cannot be simply summarized as the lower the feature dimension, the smaller the VAE loss, at least it is demonstrated that it might be a good choice to use a low-dimensional feature.

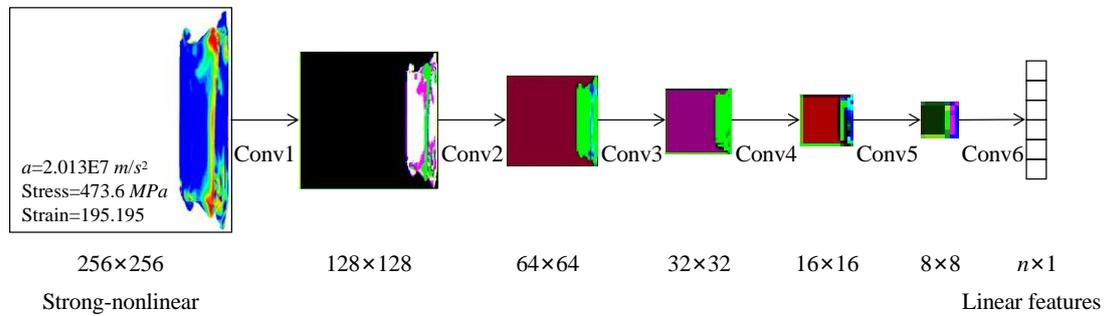

**Fig. 9.** The process of the feature extraction by the manifold learning.

---

[3] Thus, the data set contains 5,633 physical field images (training samples).

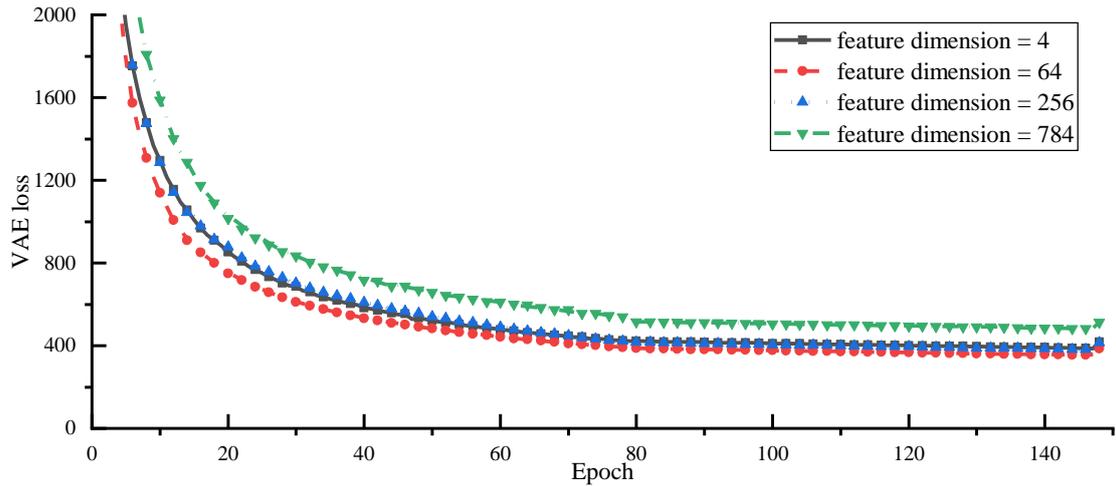

**Fig. 10.** The average VAE losses during the training processes for different feature dimensions for the cube mapping on *xOy* surface.

Meanwhile, after the dimensionality reduction, the WGAN is used to learn and generate the similar characteristic distributions of the physical field, and then the trained AVAE is applied to decode the new features. The physical fields that are generated by the CWGAN for the 4 different feature dimensions are shown in Fig. 11. No matter what the dimension of the features is, the new physical fields lack convincing details and suffer blurred regions, which make them neither realistic enough nor do they have sufficiently high resolution.

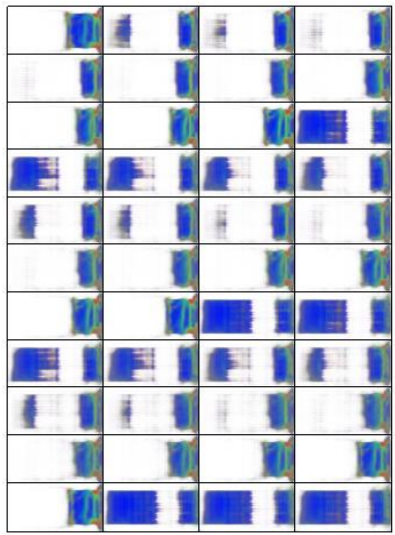
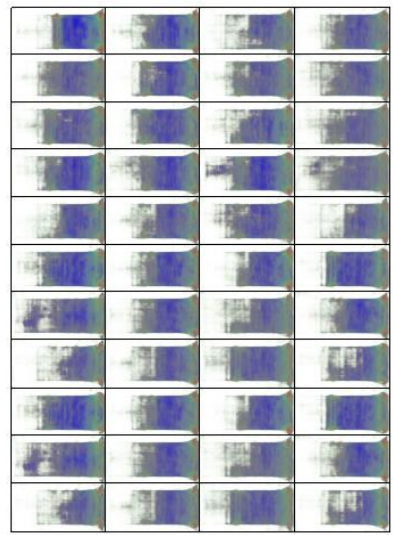

(a) Feature dimension = 4  (b) Feature dimension = 64

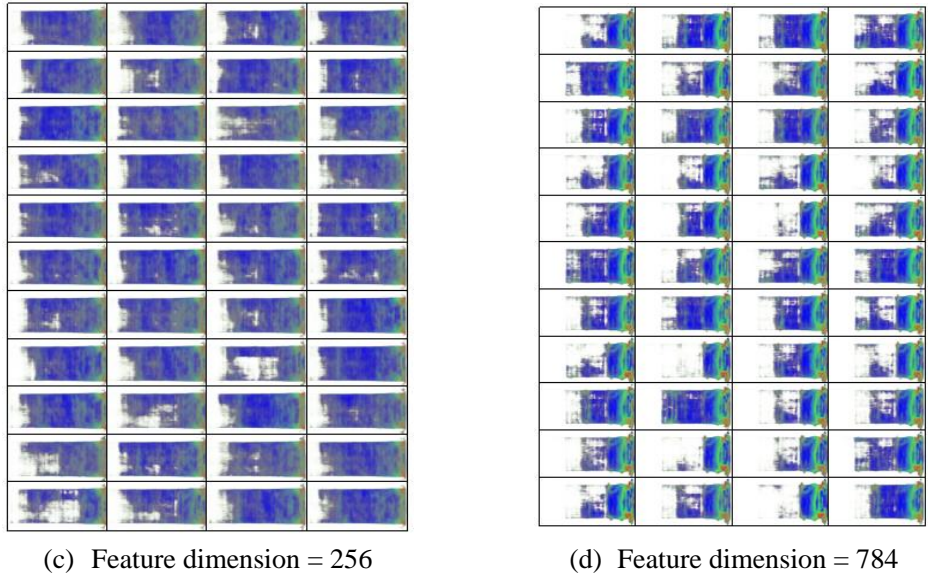

(c) Feature dimension = 256        (d) Feature dimension = 784

**Fig. 11.** Generated results by using the CWGAN in the ReConNN.

Actually, the physical quantities affecting the acceleration, stress, strain and other characteristics during the impact are all time related. Accordingly, the DOF of the nonlinear transient impact process is regarded as one, which is the time. Therefore, the dimension of the features is set to 1, and different numbers of iteration steps of the simulation of 281, 563, 1,126, 2,817 and 5,633 are run[4], respectively. The average VAE losses during the training process are shown in Fig. 12. Because this study mainly focuses on those engineering problems with sparser data or expensive computations/simulations, the lower the number of necessary training samples, the more meaningful the study. As shown in Fig. 12, it can be seen that the difference of the VAE losses between using 281 iterations and 5,633 iterations is approximately 4,000. Considering the 20-fold gap in the samples, the difference of 4,000 might be acceptable.

To further visualize the training results of the AVAE by using different samples, the decoded physical fields for several impact times are represented in Table 4. Luckily, no matter what the number of total iterations of the simulation is, the AVAE can decode features well and restore the overall structure. To better observe and compare different results, enlarged images in the 4*ms* are presented in Table 5. With the increase of the sample size, the decoded images are more similar to the actual simulation results. However, as for the results that are simulated using only 281 iterations, due to the limited sample size, the decoded image is very vague and lacks convincing details, especially in the areas that are marked by rectangles. For the results from the 5,633 iterations, the 20-fold higher number of simulation iterations truly improves the decoded results. Nevertheless, the images are still vague. Though the detailed features have been improved, there is still room for improvement. Therefore, no matter what the number of simulation iterations is, follow-up image enhancement work is necessary and this will be detailedly introduced in the following Section 4.2.3.

---

[4] In this way, 281, 563, 1,126, 2,817 and 5,633 training samples can be obtained, respectively.

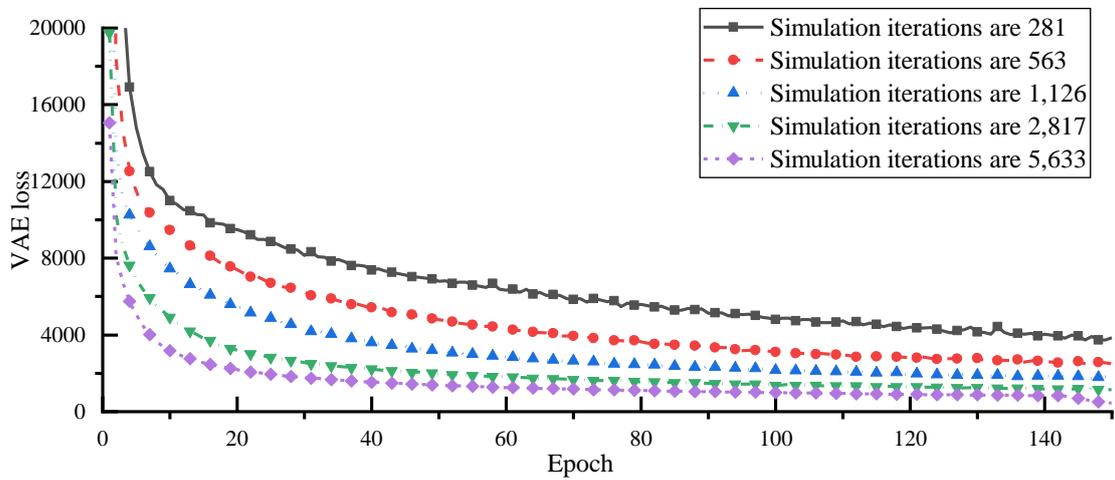

**Fig. 12.** The average VAE losses during training process when simulating different iterations for the cube mapping on *xOy* surface.

**Table 4** The decoded physical fields when simulating different iterations for the cube mapping on *xOy* surface.

where the units of the *a* and stress are $m/s^2$ and *MPa*, respectively.

| Time | Results | 281 iterations | 563 iterations | 1,126 iterations | 2,817 iterations | 5,633 iterations |
|---|---|---|---|---|---|---|
| 0.01*ms* | *a*=0<br>stress=0<br>strain=0 | | | | | |
| 0.15*ms* | *a*=3.5E+5<br>stress=1.2E-4<br>strain=1.157 | | | | | |
| 0.50*ms* | *a*=2.6E+6<br>stress=473.6<br>strain=1.840 | | | | | |

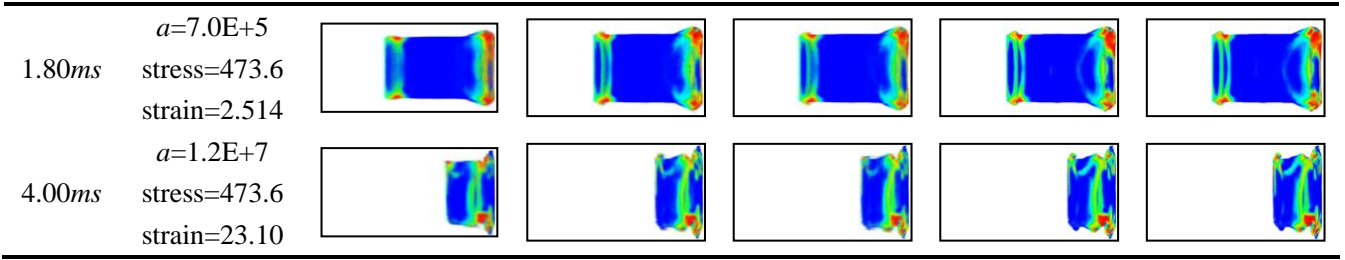

| 1.80*ms* | *a*=7.0E+5<br>stress=473.6<br>strain=2.514 | | | | | |
|---|---|---|---|---|---|---|
| 4.00*ms* | *a*=1.2E+7<br>stress=473.6<br>strain=23.10 | | | | | |

**Table 5** Enlarged images in the 4*ms* when simulating different iterations for the cube mapping on *xOy* surface.

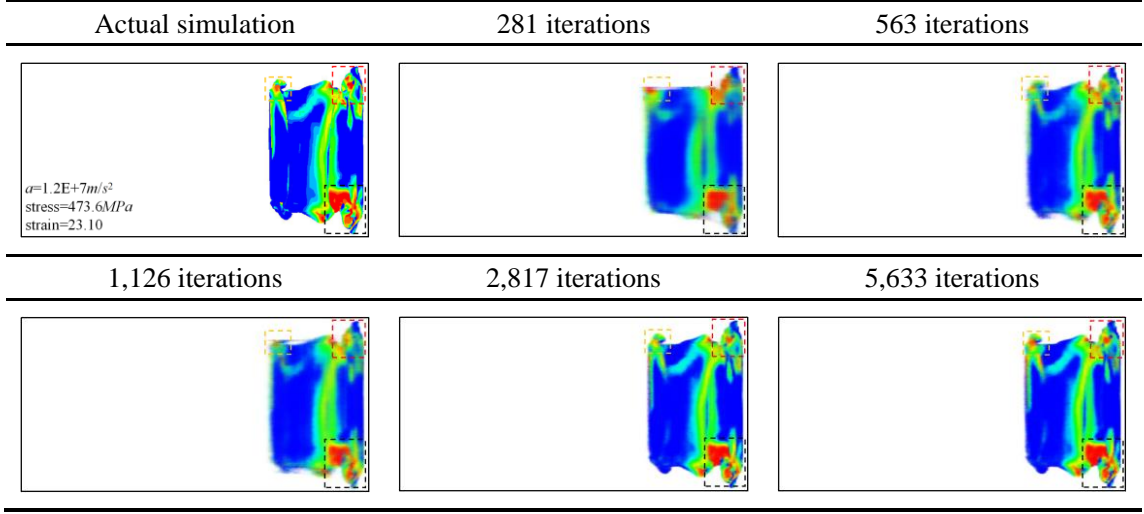

| Actual simulation | 281 iterations | 563 iterations |
|---|---|---|
| 1,126 iterations | 2,817 iterations | 5,633 iterations |

In the following step, the time-dependent ordered feature values of those cases whose feature dimension is 1 are drawn in the *xOy* coordinate, as shown in Fig. 13. Interestingly, whatever the number of simulation iterations is, the extracted features of each case can be well distributed on a smooth curve, which can further illustrate the manifold learning of the AVAE. Thus, the AVAE can well extract the features of the physical field with few simulation iterations.

In summary, to easy observe the distribution of the extracted features, and in order to conduct more convenient subsequent feature interpolation, the dimension of each feature is set as 1. Moreover, by comprehensively considering the visual effect and focusing on problems with few simulation iterations, the simulation with 563 iterations, namely, 563 samples, is selected for follow-up studies. By the way, compared with the 22,000 samples in Ref. [1] and the 6,055 samples in Ref. [2], the necessary sample size of 563 is significantly smaller, which makes this study more meaningful.

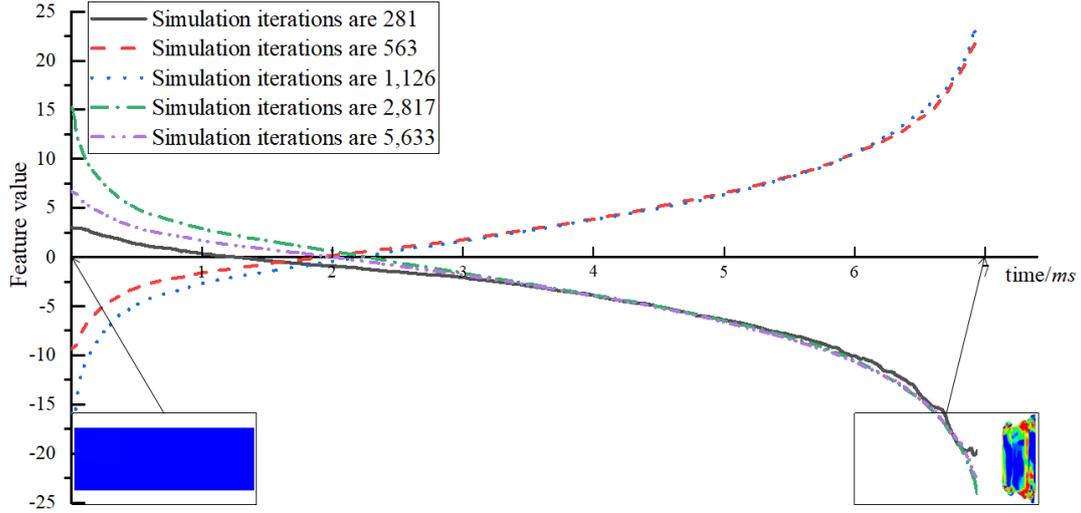

**Fig. 13.** The features of the physical field in the 1-dimensional manifold space for the cube mapping on *xOy* surface.

*4.2.2. Reconstruction of the physical field.*

In the ReConNN, after constructing the mapping from the images to the objective functions using the CIC and generating sufficient physical field images using the CWGAN, the curve of the objective functions from the simulation is interpolated. Then, the new objective functions need to be manually matched with the generated images. Hence, new images can be interpolated into the initial simulations. However, as mentioned in Section 1, the interpolation mode of the ReConNN has some shortcomings. Moreover, it is impossible to guarantee that all structures of the generated physical field images are scientific and reasonable, and so many computational resources are consumed to generate meaningless images, and distinguishing available images from unavailable ones is also inefficient. Finally, the matching work between the interpolated objective functions and the new physical field images is time-consuming and laborious.

Similar to the ReConNN, the Lagrange Interpolation (LI) is also employed in the EReConNN. However, the LI is no longer applied to the objective functions but is applied to the time-dependent ordered 1-dimensional features of the physical field. Meanwhile, the generative model is also no longer the GAN but is now the decoder of the AVAE. While training the AVAE, the feature extraction model and generative model are trained simultaneously. The LI can be expressed by

$$f(x) = \sum_{i=1}^{n} y_i p_i(x), \quad i = 1, 2, \ldots, n \tag{38}$$

s.t.

$$p_i = \prod_{i=1, i \neq j}^{n} \frac{x - x_i}{x_j - x_i} = \frac{(x - x_1)(x - x_2)\ldots(x - x_n)}{(x_j - x_1)\ldots(x_j - x_{j-1})(x_j - x_{j+1})\ldots(x_j - x_n)} \tag{39}$$

In this study, the interval of interpolation between adjacent impact iterations is 0.1. Namely, each adjacent iteration will be interpolated by 9 new values. The partial reconstructed results are shown in Table 6, where $I_i$ represents the $i$-th ($i=1, 2, \ldots, 563$) iteration that is run by the simulation, $D_j$ ($j=1, 2, \ldots, 9$) is the $j$-th interpolated step, S represents the stress whose actual number of simulation iterations is set as 5,067 ($=563 \times 9$), D-S is the interpolated results of the stress between $I_i$ and $I_{i+1}$, and Error is the error of S and D-S. Furthermore, the parts marked with red are the actual iterations that are run by simulation, while others are the interpolated results. Two iteration processes during the impact are selected to be shown: the first (from $I_7$ to $I_8$) is the period of violent impact, and the other (from $I_{324}$ to $I_{325}$) gradually converges. It can be inferred that through the decoder of the AVAE, the interpolated features are well decoded and orderly, and it is easy to guarantee that the interpolated objective functions and generated images match well. Moreover, the errors between the interpolated and actual simulation results are less than $10^{-4}$, which is sufficiently satisfactory.

**Table 6** The reconstruction of the nonlinear physical field of the impact problem for the cube mapping on $xOy$ surface.

|         | Legend | $I_7$   | $D_1$   | $D_2$   | $D_3$   | $D_4$   |
|---------|--------|---------|---------|---------|---------|---------|
| Image   | 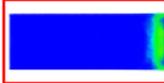 | 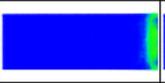 | | | | |
| Feature |        | -7.410  | -7.379  | -7.348  | -7.317  | -7.287  |
| S       |        | 404.614 | 404.972 | 405.318 | 405.650 | 405.964 |
| D-S     |        | --      | 404.959 | 405.279 | 405.580 | 405.864 |
| Error   |        | --      | 0.00%   | 0.01%   | 0.02%   | 0.02%   |
|         | $D_5$  | $D_6$   | $D_7$   | $D_8$   | $D_9$   | $I_8$   |

| Image |  |  |  |  |  |  |
|---|---|---|---|---|---|---|
| Feature | -7.257 | -7.229 | -7.201 | -7.174 | -7.149 | **-7.125** |
| S | **406.256** | **406.536** | **406.797** | **407.037** | **407.261** | **407.474** |
| D-S | 406.137 | 406.403 | 406.665 | 406.928 | 407.196 | -- |
| Error | 0.03% | 0.03% | 0.03% | 0.03% | 0.02% | -- |
| Legend | $I_{324}$ | $D_1$ | $D_2$ | $D_3$ | $D_4$ |
| Image |  |  |  |  |  |  |
| Feature |  | **3.924** | 3.926 | 3.929 | 3.931 | 3.934 |
| S |  | **473.602** | 473.603 | 473.605 | 473.608 | 473.611 |
| D-S |  | -- | 473.603 | 473.605 | 473.608 | 473.611 |
| Error |  | -- | 0.00% | 0.00% | 0.00% | 0.00% |
| $D_5$ | $D_6$ | $D_7$ | $D_8$ | $D_9$ | $I_{325}$ |
| Image |  |  |  |  |  |  |
| Feature | 3.936 | 3.939 | 3.941 | 3.944 | 3.946 | **3.949** |
| S | **473.615** | **473.619** | **473.622** | **473.625** | **473.627** | **473.627** |
| D-S | 473.615 | 473.619 | 473.622 | 473.625 | 473.627 | -- |
| Error | 0.00% | 0.00% | 0.00% | 0.00% | 0.00% | -- |

*4.2.3. Visualization Enhancement of the physical field.*

As mentioned before, the reconstructed results lack some detailed features. Therefore, it is necessary to enhance the visualization.

The decoded images are used as the inputs to the CGAN while the corresponding simulated images are the learning objectives. The enhanced results are shown in Fig. 14. For the PSNR and SSIM, empirically, if the PSNR and SSIM are larger than 20 dB and 0.9, respectively, the results are acceptable. It can be seen that the CGAN satisfactorily improves the decoded images, especially in those marked areas, and the representation of the detailed features is also greatly improved.

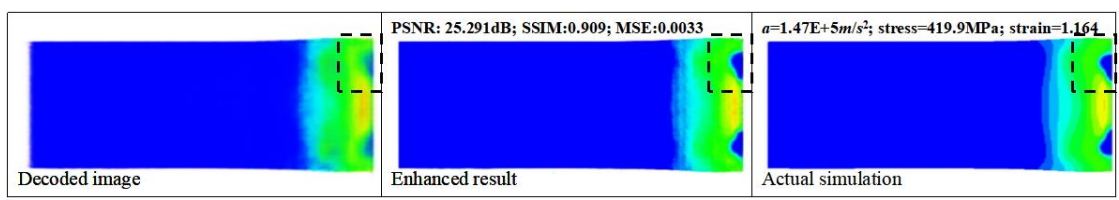

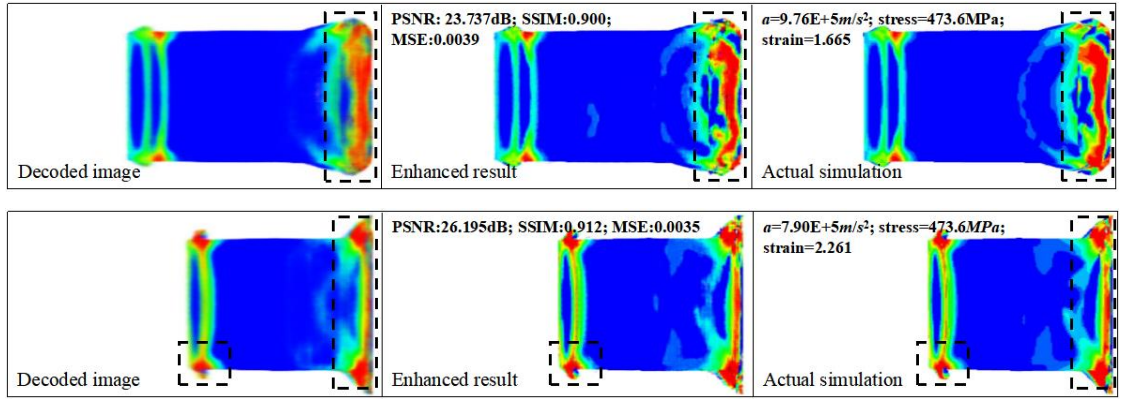

**Fig. 14.** Enhanced results of the reconstruction for the cube mapping on *xOy* surface.

*4.3. The reconstruction of the mapping on the x=y=z surface*

In this section, the impact mapping on the *x=y=z* surface is constructed. From this surface, the 3D changes and folding of the body during the impact can be better observed.

*4.3.1. Feature extraction of the physical field.*

Firstly, the AVAE is applied to extract the physical field features. As mentioned in Section 4.2.1, the final number of simulation iterations is set as 563, and so the sample size is 563. As shown in Fig. 15, after 150 training epochs, the images that are decided by the AVAE can represent the overall characteristic of the physical field. Furthermore, as shown in Fig. 16, through manifold learning, the physical field is well mapped from high-dimensional to 1-dimensional spaces.

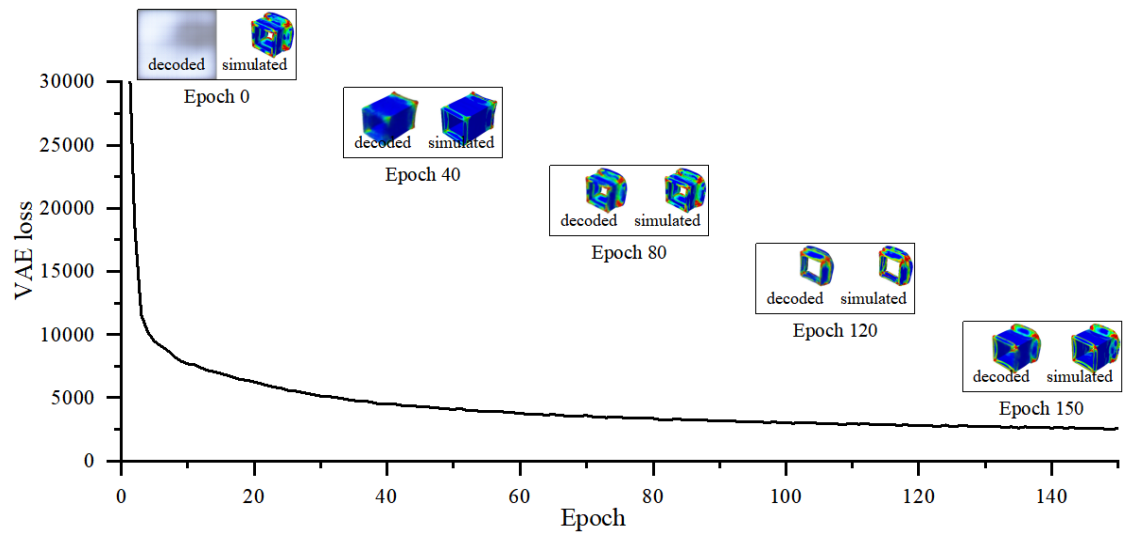

**Fig. 15.** The feature extraction process of the physical field by using the AVAE for the cube mapping on *x=y=z* surface.

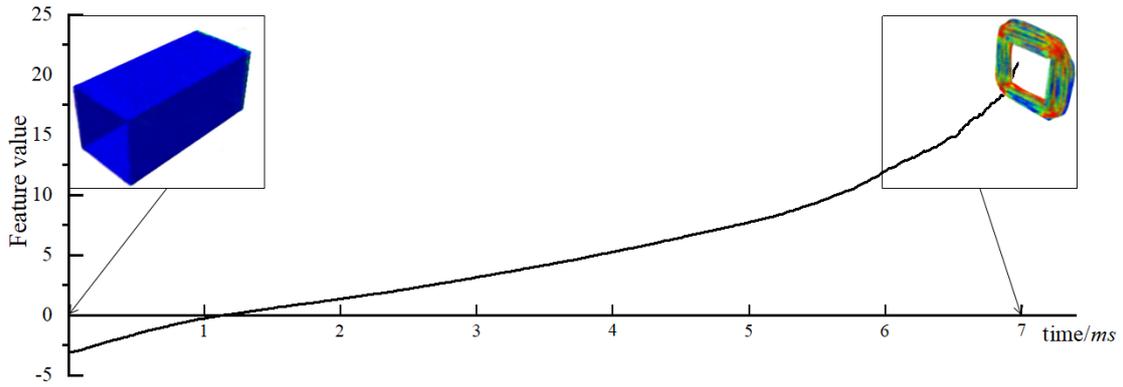

**Fig. 16.** The manifold distribution of the physical field in 1-dimensional space for the cube mapping on *x=y=z* surface.

### 4.3.2. Reconstruction of the physical field.

Identical to the reconstruction of the mapping on the *xOy* surface, the reconstructions from the 7$^{th}$ to 8$^{th}$ iterations and from the 324$^{th}$ to 325$^{th}$ iterations are shown in Table 7, respectively. Compared with the simulation process, the number of simulation iterations of the physical field easily increases by 9 times, and the added steps have high enough resolutions and many sufficiently detailed features.

**Table 7** The reconstruction of the nonlinear physical field of the impact problem for the cube mapping on *x=y=z* surface.

| | Legend | **I$_7$** | D$_1$ | D$_2$ | D$_3$ | D$_4$ |
|---|---|---|---|---|---|---|
| Image | S, Mises SNEG, (fraction = -1.0) (Avg: 75%) +4.736e+02 +4.591e+02 +4.447e+02 +4.302e+02 +4.157e+02 +4.013e+02 +3.868e+02 +3.723e+02 +3.579e+02 +3.434e+02 +3.289e+02 +3.145e+02 +3.000e+02 +0.000e+00 | | | | | |
| Feature | | **-7.410** | -7.379 | -7.348 | -7.317 | -7.287 |
| S | | **404.614** | **404.972** | **405.318** | **405.650** | **405.964** |
| D-S | | -- | 404.959 | 405.279 | 405.580 | 405.864 |
| Error | | -- | 0.00% | 0.01% | 0.02% | 0.02% |
| | D$_5$ | D$_6$ | D$_7$ | D$_8$ | D$_9$ | **I$_8$** |
| Image | | | | | | |
| Feature | -7.257 | -7.229 | -7.201 | -7.174 | -7.149 | **-7.125** |
| S | **406.256** | **406.536** | **406.797** | **407.037** | **407.261** | **407.474** |
| D-S | 406.137 | 406.403 | 406.665 | 406.928 | 407.196 | -- |
| Error | 0.03% | 0.03% | 0.03% | 0.03% | 0.02% | -- |

| | Legend | $I_{324}$ | $D_1$ | $D_2$ | $D_3$ | $D_4$ |
|---|---|---|---|---|---|---|
| Image | 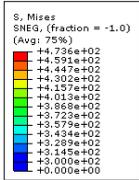 | | | | | |
| Feature | | **3.924** | 3.926 | 3.929 | 3.931 | 3.934 |
| S | | **473.602** | **473.603** | **473.605** | **473.608** | **473.611** |
| D-S | | -- | 473.603 | 473.605 | 473.608 | 473.611 |
| Error | | -- | 0.00% | 0.00% | 0.00% | 0.00% |
| | $D_5$ | $D_6$ | $D_7$ | $D_8$ | $D_9$ | $I_{325}$ |
| Image | | | | | | |
| Feature | 3.936 | 3.939 | 3.941 | 3.944 | 3.946 | **3.949** |
| S | **473.615** | **473.619** | **473.622** | **473.625** | **473.627** | **473.627** |
| D-S | 473.615 | 473.619 | 473.622 | 473.625 | 473.627 | -- |
| Error | 0.00% | 0.00% | 0.00% | 0.00% | 0.00% | -- |

### 4.3.3. Visualization enhancement of the physical field.

The enhanced results after 150 training epochs are shown in Fig. 17. Compared with the mapping on *xOy* surface, the PSNR and SSIM are slightly worse due to containing more detailed features, while the folding is better. Comprehensively, the enhanced results are satisfactory and acceptable.

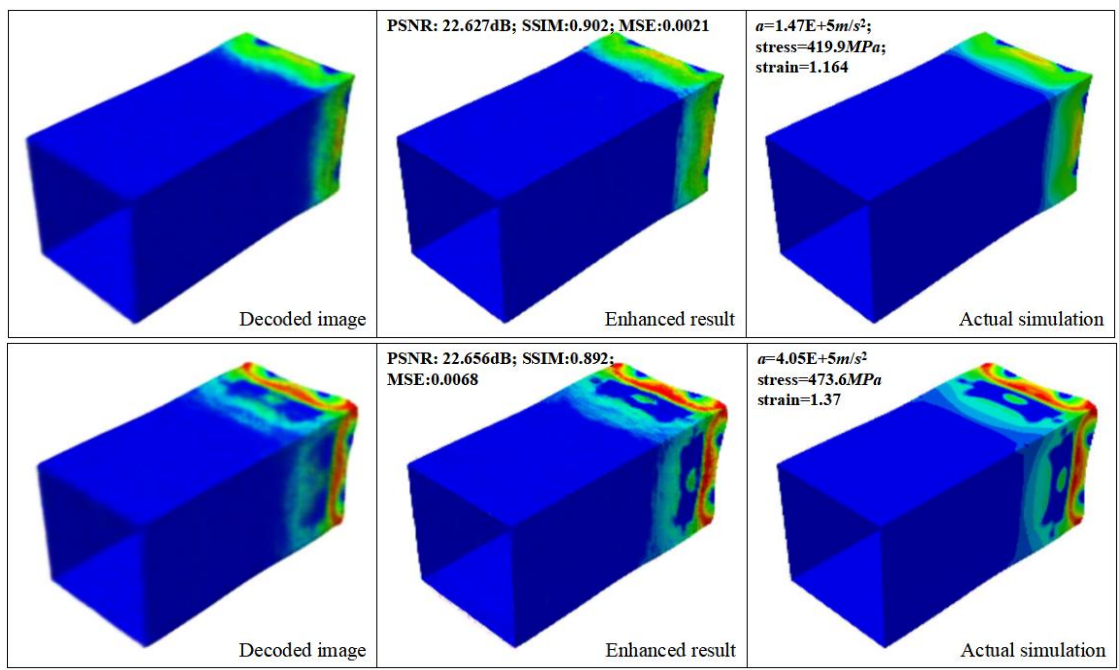

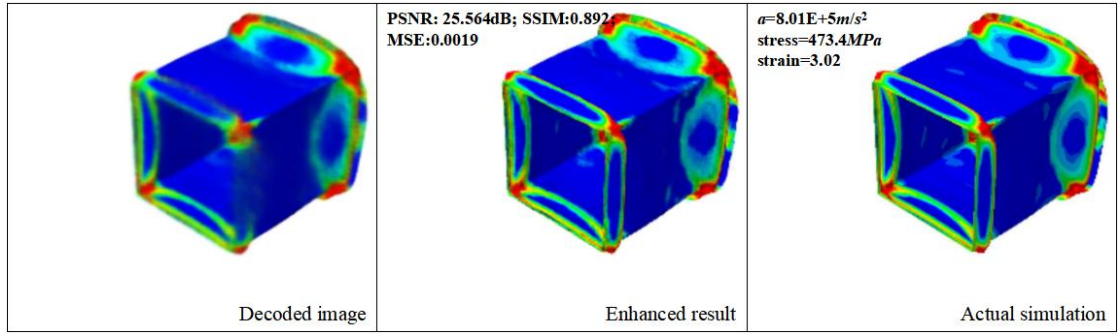

**Fig. 17.** Enhanced results of the reconstruction for the cube mapping on $x=y=z$ surface.

## 5. Experiment

After the simulated cases, an experiment is necessary to validate the feasibility of the proposed EReConNN for engineering problems. Therefore, an impact experiment is reconstructed.

Recently, the structural impacts of vehicles have attracted increasingly attention. A thin-walled metal structure is considered as a promising energy absorber due to its efficient energy absorption performance. In this section, the impact of a combined five-cell thin-walled structure that is used in a high speed train [51] is employed and simulated. Then, the impact process is reconstructed by using limited numbers of simulation iterations. Consequently, a full scale impact experiment is done to validate the reconstructed results.

*5.1. Physical model*

*5.1.1. Combined five-cell structure.*

As shown in Fig. 18, the combined multicell thin-walled aluminum structure is installed in the front end of certain high speed trains. The units in Fig. 18 are in *mm*, and the thickness is set as 5 *mm*. The material structure uses Al alloy 6,008, as shown in Table 8. It contains one octagonal and four hexagonal tubes.

**Table 8** The material parameters of the Al alloy 6,008.

| Parameter | Value |
| --- | --- |
| Young's modulus (*MPa*) | 72,000 |
| Poisson's ratio | 0.33 |
| Yield stress (*MPa*) | 131.82 |
| Density (*t*/*mm*$^3$) | $2.7 \times 10^{-9}$ |

| | Yield stress (*MPa*) | Plastic deformation |
|---|---|---|
| | 131.8 | 0 |
| | 140.0 | 0.0023 |
| | 149.4 | 0.0067 |
| | 162.3 | 0.0153 |
| | 171.2 | 0.0232 |
| Hardening curve | 178.3 | 0.0314 |
| | 183.7 | 0.0406 |
| | 187.4 | 0.0503 |
| | 189.8 | 0.0601 |
| | 190.6 | 0.0648 |
| | 191.0 | 0.0658 |

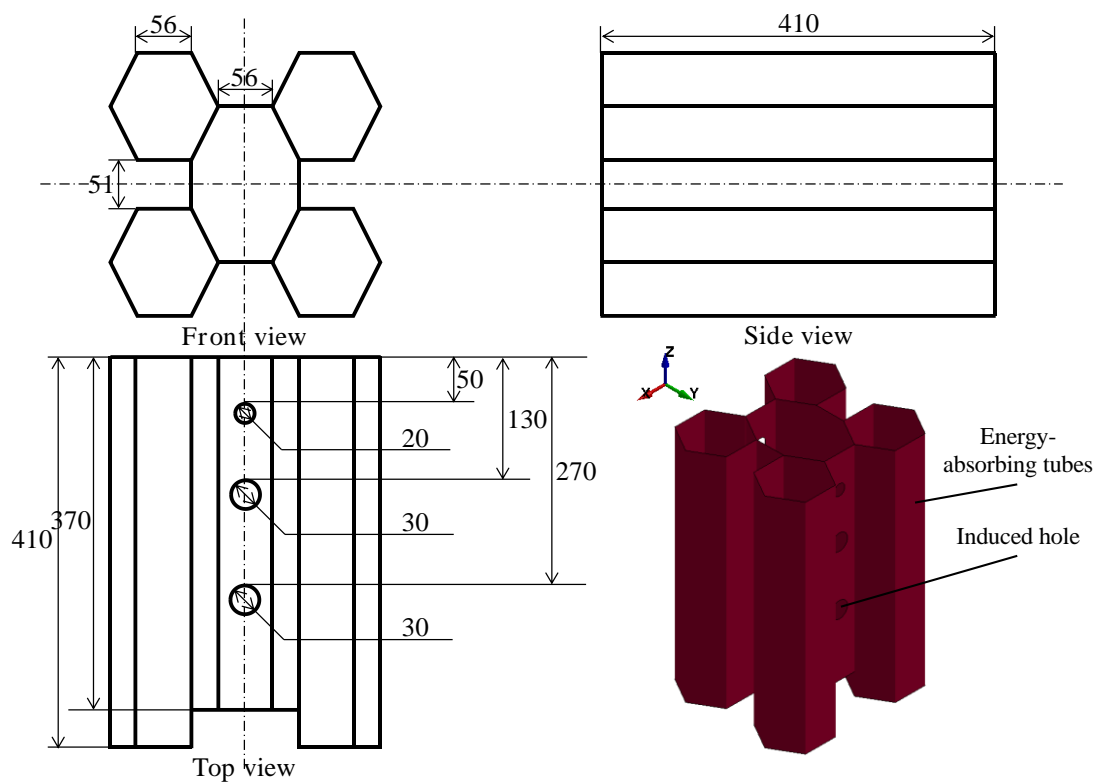

**Fig. 18.** The CAD model of the combined five-cell.

### 5.1.2. The simulation model of the impact.

As shown in Fig. 19, the impact model is composed of an impact trolley, an energy-absorbing structure and rigid tracks. To remain consistent with actual conditions, 9.81 $m/s^2$ gravity acceleration is adopted for the entire system. The frictional coefficients for the static and dynamic conditions are designed as 0.3 and 0.1, respectively. The 2,000 *kg* trolley collides at an initial velocity of 15.510 *m/s*.

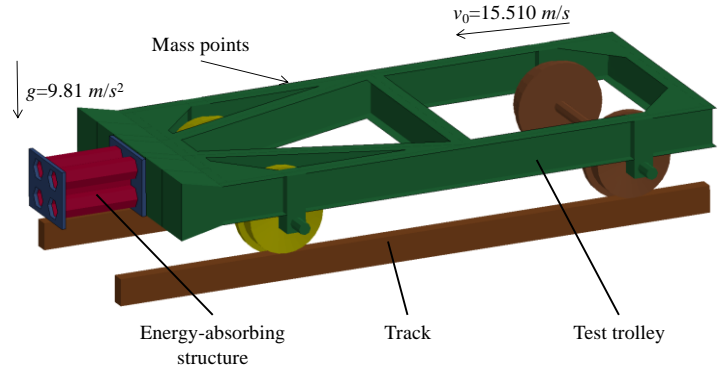

**Fig. 19.** The FE model and simulation conditions.

*5.2. Reconstruction by using the EReConNN*

In this section, the impact process mapping on the top view is reconstructed, and 100 iterations are simulated to obtain training samples.

*5.2.1. Feature extraction of the physical field.*

The manifold distribution of the physical field in a 1-dimensional space is represented in Figs. 20. It can be found that the AVAE well maps the physical field of the combined five-cell structure from high-dimensional to 1-dimensional manifold spaces.

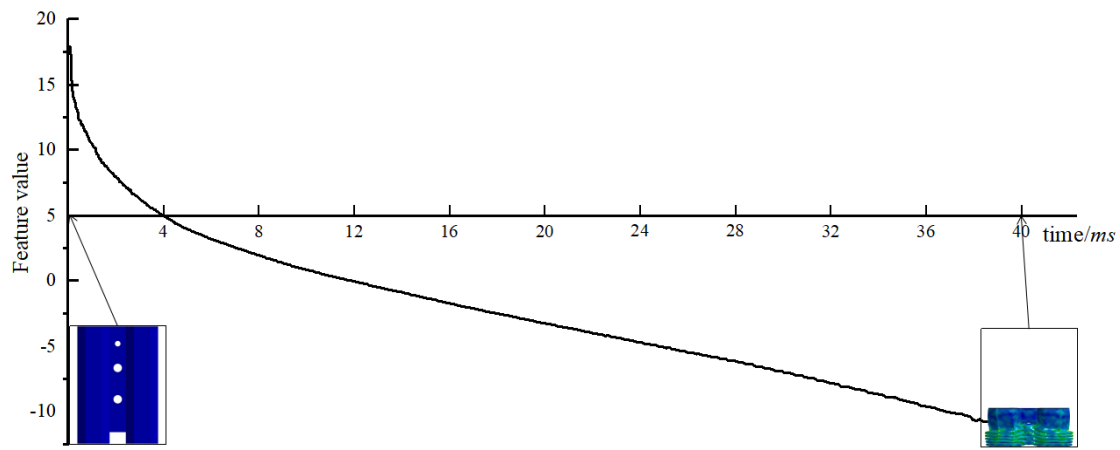

**Fig. 20.** The manifold distribution in 1-dimensional space for the combined five-cell.

*5.2.2. Reconstruction of the physical field.*

The impact process takes 40 *ms*, and the impact images at 0 *ms*, 0.4 *ms*, 0.8 *ms*, …, 40 *ms* in simulation are saved. Namely, the images are sampled at 2,500 frames per second. As shown in Table 9, similar to the reconstruction in Section 4 and taking 0.04 *ms* as the interpolation step, 9 times the number of iterations of the initial simulation is obtained, namely, the images are sampled at 25,000 frames per second. Furthermore, an experiment is done to evaluate the reconstruction results. The impact process of the experiment is captured by a high speed camera with a frequency of 5,000 frames per second.

It can be seen that the reconstructed impact process is consistent with the experiment. The tubes buckle symmetrically along the central axis. Moreover, a 5 times (if necessary, more times is feasible) greater sampling frequency is adequately used compared with the high speed camera, which can help us to effectively reduce the computational and equipment costs.

**Table 9** Comparisons of impact series between reconstructed and experimental.

| Time | Reconstruction (25,000 frame/*s*) | Experiment (5,000 frame/*s*) |
| --- | --- | --- |
| 0.6*ms* | 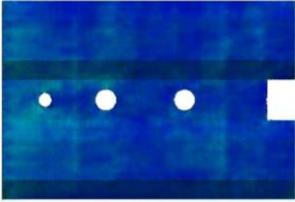 | 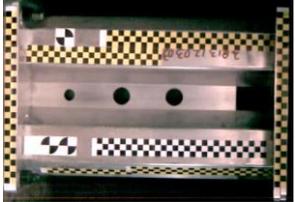 |
| 8.2*ms* | 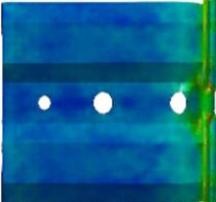 | 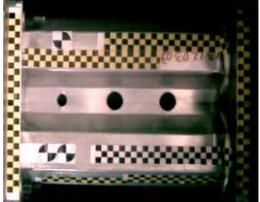 |
| 16.2*ms* | 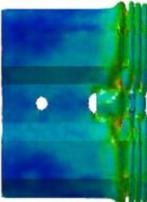 | 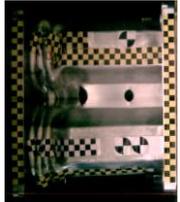 |
| 24.2*ms* | 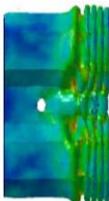 | 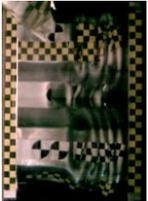 |
| 32.2*ms* | 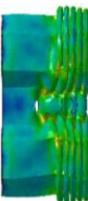 | 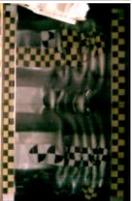 |

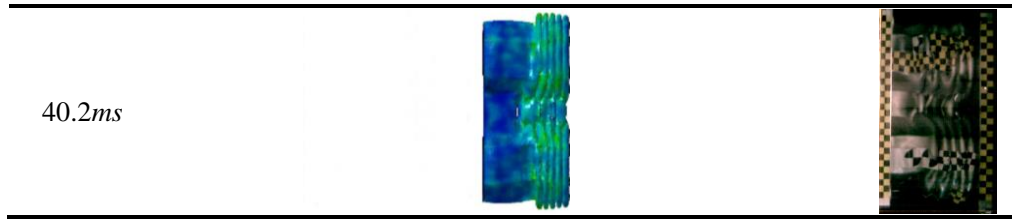

*40.2ms*

## Conclusions

In this study, an EReConNN is developed to solve a nonlinear transient impact case. Simultaneously, the EReConNN addresses some shortcomings of the existing ReConNN. The contributions of this study can be summarized as follows.

i. The proposed EReConNN is successfully applied to the nonlinear transient case of an impact problem.

ii. The adversarial algorithm integrated with VAE and an AVAE is proposed.

iii. The integrated mode of the CNN and GAN is replaced by the proposed AVAE, which completes the feature extraction and time-dependent ordered image generation simultaneously.

iv. After reconstruction, the CGAN is innovatively employed to perform the image postprocessing. This makes the reconstructive results more meaningful and reasonable.

v. Finally, an engineering problem is reconstructed and experimented. The results present that the proposed EReConNN can effectively reduce the computational and equipment costs.

## Acknowledgments

This work has been supported by Project of the Key Program of National Natural Science Foundation of China under the Grant Numbers 11572120 and 51621004, Key Projects of the Research Foundation of Education Bureau of Hunan Province (17A224).